\documentclass[superscriptaddress,preprintnumbers,showpacs,amsmath,amssymb,prl,twocolumn]{revtex4}

\usepackage{graphicx}
\usepackage{dcolumn}
\usepackage{bm}
\usepackage{ifthen}
\usepackage{setspace}
\usepackage{color}

\def\ra{\rightarrow}
\newcommand{\GeV}{\ensuremath{\mathrm{Ge\kern -0.1em V}}}
\newcommand{\MeV}{\ensuremath{\mathrm{Me\kern -0.1em V}}}

\newcommand{\fb}{\ensuremath{\mathrm{fb}^{-1}}}

\newcommand{\mm}{\ensuremath{\mu^{+}\mu^{-}}}

\newcommand{\iso}{\ensuremath{\mathit{I}}}

\def \ptB {p_{T}(B)}

\newcommand{\cdf}{CDF~II}

\def \bsdecay  {B^0_s \rightarrow e^{+}\mu^{-}}
\def \bseldecay  {B^0_s \rightarrow e^{+}\ell^{-}}
\def \bddecay  {B^0 \rightarrow e^{+}\mu^{-}}

\def \bstoeedecay  {B^0_s \rightarrow e^{+}e^{-}}
\def \bdtoeedecay  {B^0 \rightarrow e^{+}e^{-}}

\def \brs    {\mathcal{B}(\bsdecay)}
\def \brsel  {\mathcal{B}(\bseldecay)}
\def \brd    {\mathcal{B}(\bddecay)}

\def \brstoee    {\mathcal{B}(\bstoeedecay)}
\def \brdtoee    {\mathcal{B}(\bdtoeedecay)}

\def \brslimit {\brs <  2.0~(2.6) \times 10^{-7}}
\def \brdlimit {\brd <  6.4~(7.9) \times 10^{-8}}
\def \psslimit { { M_{LQ}}(B^0_s\rightarrow e^+ \mu^-) >  47.8~(44.9) \; {\rm TeV/c^2}}
\def \psdlimit { { M_{LQ}}(B^0\rightarrow e^+ \mu^-) >  59.3~(56.3) \; {\rm TeV/c^2}}

\def \brstoeelimit {\brstoee <  2.8 \times 10^{-7}}
\def \brdtoeelimit {\brdtoee <  8.3 \times 10^{-8}}

\def \brslimitnt {\brs <  2.0 \times 10^{-7}}
\def \brdlimitnt {\brd <  6.4 \times 10^{-8}}
\def \psslimitnt { { M_{LQ}}(B^0_s\rightarrow e^+ \mu^-) >  47.8 \; {\rm TeV/c^2}}
\def \psdlimitnt { { M_{LQ}}(B^0\rightarrow e^+ \mu^-) >  59.3 \; {\rm TeV/c^2}}

\def \brstoeelimitnt {\brstoee <  2.8 \times 10^{-7}}
\def \brdtoeelimitnt {\brdtoee <  8.3 \times 10^{-8}}

\def \mm  {\mu^{+}\mu^{-}}

\def \bdkpdecay {B^0 \rightarrow K^{+}\pi^{-}}
\def \brbdkp   {\mathcal{B}( \bdkpdecay )}

\begin{document}

\title{  Search for the Decays ${ B^0_{(s)}\rightarrow e^+\mu^-}$  and ${ B^0_{(s)}\rightarrow e^+e^-}$ in CDF Run II }

\begin{abstract}
We report results from a search for the lepton flavor violating decays  $B^0_{(s)}\rightarrow e^+\mu^-$, 
and the flavor-changing neutral-current decays $B^0_{(s)} \rightarrow e^+ e^-$. The analysis uses
data corresponding to ${\rm 2\; fb^{-1}}$   of integrated luminosity  of $p \bar{p}$   collisions at
$\sqrt{s}=1.96~{\rm TeV}$
collected with the upgraded Collider  Detector (CDF II) at the Fermilab Tevatron.
The  observed number of  $B^0_{(s)}$ candidates is consistent with background expectations.
The resulting Bayesian upper limits on the  branching ratios  at 90\% credibility level are $\brslimitnt$,  $\brdlimitnt$, 
$\brstoeelimitnt $ and $ \brdtoeelimitnt $.  
From the limits on $\mathcal{B} (B^0_{(s)}\rightarrow e^+\mu^-)$, the following lower bounds on 
the Pati-Salam leptoquark masses are also derived:  $\psslimitnt$, and  $\psdlimitnt$,  at 90\% credibility level.

\end{abstract}

\affiliation{Institute of Physics, Academia Sinica, Taipei, Taiwan 11529, Republic of China} 
\affiliation{Argonne National Laboratory, Argonne, Illinois 60439} 
\affiliation{University of Athens, 157 71 Athens, Greece} 
\affiliation{Institut de Fisica d'Altes Energies, Universitat Autonoma de Barcelona, E-08193, Bellaterra (Barcelona), Spain} 
\affiliation{Baylor University, Waco, Texas  76798} 
\affiliation{Istituto Nazionale di Fisica Nucleare Bologna, $^x$University of Bologna, I-40127 Bologna, Italy} 
\affiliation{Brandeis University, Waltham, Massachusetts 02254} 
\affiliation{University of California, Davis, Davis, California  95616} 
\affiliation{University of California, Los Angeles, Los Angeles, California  90024} 
\affiliation{University of California, San Diego, La Jolla, California  92093} 
\affiliation{University of California, Santa Barbara, Santa Barbara, California 93106} 
\affiliation{Instituto de Fisica de Cantabria, CSIC-University of Cantabria, 39005 Santander, Spain} 
\affiliation{Carnegie Mellon University, Pittsburgh, PA  15213} 
\affiliation{Enrico Fermi Institute, University of Chicago, Chicago, Illinois 60637}
\affiliation{Comenius University, 842 48 Bratislava, Slovakia; Institute of Experimental Physics, 040 01 Kosice, Slovakia} 
\affiliation{Joint Institute for Nuclear Research, RU-141980 Dubna, Russia} 
\affiliation{Duke University, Durham, North Carolina  27708} 
\affiliation{Fermi National Accelerator Laboratory, Batavia, Illinois 60510} 
\affiliation{University of Florida, Gainesville, Florida  32611} 
\affiliation{Laboratori Nazionali di Frascati, Istituto Nazionale di Fisica Nucleare, I-00044 Frascati, Italy} 
\affiliation{University of Geneva, CH-1211 Geneva 4, Switzerland} 
\affiliation{Glasgow University, Glasgow G12 8QQ, United Kingdom} 
\affiliation{Harvard University, Cambridge, Massachusetts 02138} 
\affiliation{Division of High Energy Physics, Department of Physics, University of Helsinki and Helsinki Institute of Physics, FIN-00014, Helsinki, Finland} 
\affiliation{University of Illinois, Urbana, Illinois 61801} 
\affiliation{The Johns Hopkins University, Baltimore, Maryland 21218} 
\affiliation{Institut f\"{u}r Experimentelle Kernphysik, Universit\"{a}t Karlsruhe, 76128 Karlsruhe, Germany} 
\affiliation{Center for High Energy Physics: Kyungpook National University, Daegu 702-701, Korea; Seoul National University, Seoul 151-742, Korea; Sungkyunkwan University, Suwon 440-746, Korea; Korea Institute of Science and Technology Information, Daejeon, 305-806, Korea; Chonnam National University, Gwangju, 500-757, Korea} 
\affiliation{Ernest Orlando Lawrence Berkeley National Laboratory, Berkeley, California 94720} 
\affiliation{University of Liverpool, Liverpool L69 7ZE, United Kingdom} 
\affiliation{University College London, London WC1E 6BT, United Kingdom} 
\affiliation{Centro de Investigaciones Energeticas Medioambientales y Tecnologicas, E-28040 Madrid, Spain} 
\affiliation{Massachusetts Institute of Technology, Cambridge, Massachusetts  02139} 
\affiliation{Institute of Particle Physics: McGill University, Montr\'{e}al, Qu\'{e}bec, Canada H3A~2T8; Simon Fraser University, Burnaby, British Columbia, Canada V5A~1S6; University of Toronto, Toronto, Ontario, Canada M5S~1A7; and TRIUMF, Vancouver, British Columbia, Canada V6T~2A3} 
\affiliation{University of Michigan, Ann Arbor, Michigan 48109} 
\affiliation{Michigan State University, East Lansing, Michigan  48824}
\affiliation{Institution for Theoretical and Experimental Physics, ITEP, Moscow 117259, Russia} 
\affiliation{University of New Mexico, Albuquerque, New Mexico 87131} 
\affiliation{Northwestern University, Evanston, Illinois  60208} 
\affiliation{The Ohio State University, Columbus, Ohio  43210} 
\affiliation{Okayama University, Okayama 700-8530, Japan} 
\affiliation{Osaka City University, Osaka 588, Japan} 
\affiliation{University of Oxford, Oxford OX1 3RH, United Kingdom} 
\affiliation{Istituto Nazionale di Fisica Nucleare, Sezione di Padova-Trento, $^y$University of Padova, I-35131 Padova, Italy} 
\affiliation{LPNHE, Universite Pierre et Marie Curie/IN2P3-CNRS, UMR7585, Paris, F-75252 France} 
\affiliation{University of Pennsylvania, Philadelphia, Pennsylvania 19104}
\affiliation{Istituto Nazionale di Fisica Nucleare Pisa, $^z$University of Pisa, $^{aa}$University of Siena and $^{bb}$Scuola Normale Superiore, I-56127 Pisa, Italy} 
\affiliation{University of Pittsburgh, Pittsburgh, Pennsylvania 15260} 
\affiliation{Purdue University, West Lafayette, Indiana 47907} 
\affiliation{University of Rochester, Rochester, New York 14627} 
\affiliation{The Rockefeller University, New York, New York 10021} 
\affiliation{Istituto Nazionale di Fisica Nucleare, Sezione di Roma 1, $^{cc}$Sapienza Universit\`{a} di Roma, I-00185 Roma, Italy} 

\affiliation{Rutgers University, Piscataway, New Jersey 08855} 
\affiliation{Texas A\&M University, College Station, Texas 77843} 
\affiliation{Istituto Nazionale di Fisica Nucleare Trieste/Udine, I-34100 Trieste, $^{dd}$University of Trieste/Udine, I-33100 Udine, Italy} 
\affiliation{University of Tsukuba, Tsukuba, Ibaraki 305, Japan} 
\affiliation{Tufts University, Medford, Massachusetts 02155} 
\affiliation{Waseda University, Tokyo 169, Japan} 
\affiliation{Wayne State University, Detroit, Michigan  48201} 
\affiliation{University of Wisconsin, Madison, Wisconsin 53706} 
\affiliation{Yale University, New Haven, Connecticut 06520} 
\author{T.~Aaltonen}
\affiliation{Division of High Energy Physics, Department of Physics, University of Helsinki and Helsinki Institute of Physics, FIN-00014, Helsinki, Finland}
\author{J.~Adelman}
\affiliation{Enrico Fermi Institute, University of Chicago, Chicago, Illinois 60637}
\author{T.~Akimoto}
\affiliation{University of Tsukuba, Tsukuba, Ibaraki 305, Japan}
\author{B.~\'{A}lvarez~Gonz\'{a}lez$^s$}
\affiliation{Instituto de Fisica de Cantabria, CSIC-University of Cantabria, 39005 Santander, Spain}
\author{S.~Amerio$^y$}
\affiliation{Istituto Nazionale di Fisica Nucleare, Sezione di Padova-Trento, $^y$University of Padova, I-35131 Padova, Italy} 

\author{D.~Amidei}
\affiliation{University of Michigan, Ann Arbor, Michigan 48109}
\author{A.~Anastassov}
\affiliation{Northwestern University, Evanston, Illinois  60208}
\author{A.~Annovi}
\affiliation{Laboratori Nazionali di Frascati, Istituto Nazionale di Fisica Nucleare, I-00044 Frascati, Italy}
\author{J.~Antos}
\affiliation{Comenius University, 842 48 Bratislava, Slovakia; Institute of Experimental Physics, 040 01 Kosice, Slovakia}
\author{G.~Apollinari}
\affiliation{Fermi National Accelerator Laboratory, Batavia, Illinois 60510}
% ...

\collaboration{CDF Collaboration\footnote{With visitors from $^a$University of Massachusetts Amherst, Amherst, Massachusetts 01003,
$^b$Universiteit Antwerpen, B-2610 Antwerp, Belgium, 
$^c$University of Bristol, Bristol BS8 1TL, United Kingdom,
$^d$Chinese Academy of Sciences, Beijing 100864, China, 
$^e$Istituto Nazionale di Fisica Nucleare, Sezione di Cagliari, 09042 Monserrato (Cagliari), Italy,
$^f$University of California Irvine, Irvine, CA  92697, 
$^g$University of California Santa Cruz, Santa Cruz, CA  95064, 
$^h$Cornell University, Ithaca, NY  14853, 
$^i$University of Cyprus, Nicosia CY-1678, Cyprus, 
$^j$University College Dublin, Dublin 4, Ireland,
$^k$University of Edinburgh, Edinburgh EH9 3JZ, United Kingdom, 
$^l$University of Fukui, Fukui City, Fukui Prefecture, Japan 910-0017
$^m$Kinki University, Higashi-Osaka City, Japan 577-8502
$^n$Universidad Iberoamericana, Mexico D.F., Mexico,
$^o$Queen Mary, University of London, London, E1 4NS, England,
$^p$University of Manchester, Manchester M13 9PL, England, 
$^q$Nagasaki Institute of Applied Science, Nagasaki, Japan, 
$^r$University of Notre Dame, Notre Dame, IN 46556,
$^s$University de Oviedo, E-33007 Oviedo, Spain, 
$^t$Texas Tech University, Lubbock, TX  79609, 
$^u$IFIC(CSIC-Universitat de Valencia), 46071 Valencia, Spain,
$^v$University of Virginia, Charlottesville, VA  22904,
$^w$Bergische Universit\"at Wuppertal, 42097 Wuppertal, Germany,
$^{ee}$On leave from J.~Stefan Institute, Ljubljana, Slovenia, 
}}
\noaffiliation


\maketitle

Rare  particle decays that are either forbidden within the standard model of
particle physics (SM), or are expected to have very small branching ratios 
provide excellent signatures with which to look for new physics and allow to
probe subatomic processes that are beyond the reach of direct searches. 
The decays $B^0_{(s)}\rightarrow e^+ \mu^-$~\cite{qconj} 
are forbidden within the SM, in which lepton number and lepton flavor are
conserved. However the  observation of neutrino oscillations indicates that 
lepton flavor is not conserved. 
To date, no lepton flavor violating (LFV) decays in the charged sector 
such as $B^0_{(s)}\rightarrow e^+ \mu^-$ have been observed. 
These decays are allowed in models where the SM has  been extended by 
heavy singlet Dirac neutrinos~\cite{DIRAC}. The LFV decays are also allowed
in some physics scenarios beyond the SM,  such as the  Pati-Salam 
model~\cite{pati-salam} and supersymmetry (SUSY) models~\cite{SUSY}.
The grand-unification theory by J. Pati and A. Salam predicts a new 
interaction to mediate transitions between leptons and quarks via exchange of  
spin-1 gauge bosons, which are called Pati-Salam leptoquarks (LQ), that carry both
 color and lepton quantum numbers~\cite{pati-salam}.  
The lepton and quark components of the leptoquarks are not 
necessarily from the same generation~\cite{scott,Blanke}, and  
the decays  $B^0_{s}\rightarrow e^+ \mu^-$ and $B^0\rightarrow e^+ \mu^-$
can be mediated by different types of leptoquarks. 
Processes involving flavor-changing neutral currents (FCNCs)
can occur in the SM only through higher-order Feynman diagrams 
where new physics contributions can provide a significant enhancement.
Compared to $B^0_{(s)} \rightarrow \mu^+ \mu^-$~\cite{btomumu}, the FCNC decays 
of  $ B^0_{(s)} \rightarrow e^+ e^-$ are further suppressed by the square of 
the ratio of the electron and muon masses $(m_e/m_\mu)^2$.
The SM expectations for branching ratios of  $ B^0_{(s)} \rightarrow e^+ e^-$   
are of the order of $ 10^{-15}$~\cite{Misiak}.

In this  Letter we report on a search for the LFV decays $B^0_{(s)}\rightarrow e^+\mu^-$  
and the FCNC decays $B^0_{(s)} \rightarrow e^+ e^-$,   using a data sample corresponding to  
2 $\fb$of integrated luminosity collected in  $p \bar{p}$ collisions at $\sqrt{s} = 1.96$~TeV.
With no evidence for either the LFV or  FCNC decays, we set upper limits on their 
branching ratios using the common reference decay   $B^0 \rightarrow K^+ \pi^-$, which has a 
precisely-known  branching ratio. 
This is the first time a search for  $B^0_s  \rightarrow e^+ e^-$  
has been performed.

A detailed description of the \cdf\ detector can be found in Ref.~\cite{JpsiPRD}. 
Here we give a brief description of the detector elements most relevant to 
this analysis. 
Charged particle tracking is provided by a silicon microstrip detector  
together with the surrounding  open-cell wire drift chamber (COT), 
both immersed in a 1.4 T axial magnetic field. 
The tracking system provides precise vertex and momentum 
measurement for charged particles in the pseudorapidity range 
$\left|\eta\right|<1.0$~\cite{csys}. 
Surrounding the tracking system are electromagnetic (CEM) 
  and hadronic 
 sampling calorimeters, arranged in a projective geometry.
Drift chambers and scintillation counters  
are located behind the calorimeters to detect muons within 
$\left|\eta\right|<0.6$ (CMU) and $0.6<\left|\eta\right|<1.0$ (CMX).  

We use a data sample enriched in 
two-body $B$--decays selected by a three-level trigger system using the extremely fast tracker~\cite{XFT} 
at level-1, and the 
silicon vertex trigger~\cite{SVT} at level-2. The trigger 
requires two  oppositely-charged tracks, each with a transverse momentum $p_{T}>2~{\rm GeV/c}$, 
and an impact parameter~\cite{IP}  $0.1 <  d_0 < 1$ mm.
It also requires the scalar sum of the transverse momenta of the two tracks  to be greater than 5.5 GeV/c, 
the difference in the azimuthal angles of  the tracks $20^\circ <\Delta \varphi<135^\circ$, 
and a transverse decay length~\cite{lxy}  $L_{xy}>200$ $\mu$m.
At the level-3 trigger stage, and in the offline analysis, the trigger selections are enforced with a more accurate 
determination of the same quantities. In the off-line analysis, additionally  we require: the $B$--meson  isolation  
${\rm I}>0.675$~\cite{isodef},
the  pointing angle  $\Delta \phi<6.3^\circ$~\cite{poidef}, and  
a tighter selection of $L_{xy}>375$ $\mu$m. These three thresholds were optimized  in an unbiased way to obtain the best 
sensitivity for the searches using 
the procedure described in Ref.~\cite{punzi}.  

Electron and muon identification is applied in the selection of   
${ B^0_{(s)}\rightarrow e^+\mu^-}$  and ${ B^0_{(s)}\rightarrow e^+e^-}$  decay modes.
The electron identification~\cite{eID} requires that both the specific ionization ($\mathit{dE/dx}$) 
measured in the COT,  and the transverse and longitudinal shower shape as measured in the CEM, be 
consistent with the hypothesis that the particle is an electron.   
The performance of electron identification is optimized using pure electron samples reconstructed 
from  $\gamma \rightarrow e^+e^-$ conversions and hadron and muon samples from   
$D^0 \rightarrow K^- \pi^+$, $\Lambda \rightarrow p\pi^-$, and $J/\psi\ra\mm$ decays.
We find the identification efficiency to be  around ~70\% 
for electrons.
The muon identification starts from tracks in the COT that are extrapolated 
into the muon detectors and are required to match hits in the muon systems. 
The muon selection is fully efficient for muons with  $p_T >$ 2 GeV/c in
CMU or CMX.

The mass resolution  $\sigma_{m}$ of fully-reconstructed $B$--meson
decays to two charged particles is about 28~MeV/c$^{2}$.
Energy loss due to bremsstrahlung by electrons
generates a tail on the low side of the mass distribution.
This tail is more 
prominent for the  $B^0_{(s)} \rightarrow e^+ e^-$ channels, where two electrons are involved. 
We define search windows of (5.262--5.477)~GeV/c$^2$ for  $B^0_{s} \rightarrow e^+ \mu^-$ and  
(5.171--5.387)~GeV/c$^2$  for $B^0 \rightarrow e^+ \mu^-$.   
These correspond to a window  around the nominal values of the 
$B^0_s$  and  $B^0$ masses~\cite{PDG} of approximately $\pm 3 \sigma_{m}$. 
To recover some of the acceptance loss due to electron bremsstrahlung  for the 
$B^0_{(s)} \rightarrow e^+ e^-$ channels, we choose 
wider and asymmetric search windows ranging  from 6  $\sigma_{m}$ below  to 3   $\sigma_{m}$  above the nominal values 
of the $B^0_s$  and $B^0$ masses. The search windows are 
(5.154--5.477)~GeV/c$^2$ for  the $B^0_s$ 
and (5.064--5.387)~GeV/c$^2$ for the $B^0$. The sideband regions (4.800--5.028)~GeV/c$^2$  and 
(5.549--5.800)~GeV/c$^2$  are used to estimate the combinatorial backgrounds.

The background contributions considered include combinations of
random track pairs and partial $B$ decays that accidentally meet the 
selection requirement (combinatorial),
and hadronic two-body $B$ decays in which both final
particles are misidentified as leptons. The
combinatorial background is evaluated by extrapolating the normalized
number of events found in the sidebands to the signal region. 
The double-lepton misidentification rate is determined by applying electron and muon 
misidentification probabilities to the number of two-body decays found in the
search window.

\begin{figure}[htbp]
\begin{center}
\includegraphics[width=3.5in]{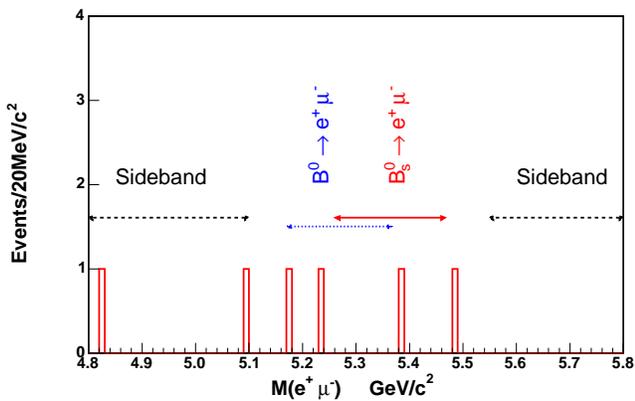}

\caption{Invariant mass distribution of $e^+ \mu^-$  pairs
for events where one track passed the electron identification and the other track the muon identification.  
The $B^0_s$ ($B^0$) search window is indicated by the solid (dotted) line.
The sideband regions are indicated by the dashed lines.}  
\label{massfinalemu}
\end{center}
\end{figure}

Figure~\ref{massfinalemu} shows  
the invariant mass distribution  for $e^+ \mu^-$ candidates. 
We  observe  one event in the  $B^0_s$ mass window, and two events in the $B^0$ mass window, consistent
with the estimated total background of  $0.8\pm 0.6$ events in  the $B^0_s$  search window, and $0.9 \pm 0.6$ in the $B^0$  window. 
The combinatorial background in both channels is estimated to be $0.7 \pm 0.6$ events. The number of events where
two tracks are 
misidentified as  electron and muon is estimated to be $0.09 \pm 0.02$ for the $B^0_s$ case and  $0.22 \pm 0.04$  for the $B^0$ case.
 
\begin{figure}[htbp]
\begin{center}
\includegraphics[width=3.5in]{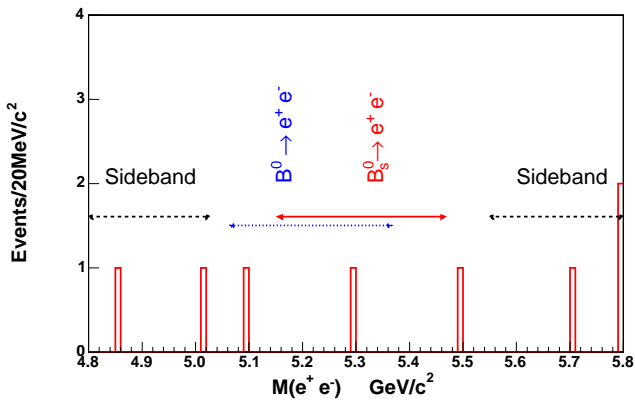}
\caption{  Invariant mass distributions of  $e^+ e^-$ pairs
for events where both tracks passed the electron identification.
The $B^0_s$ ($B^0$) search window is indicated by the solid (dotted) line.
The sideband regions are indicated by dashed lines.}
\label{massfinalee}
\end{center}
\end{figure}

Figure~\ref{massfinalee} shows the invariant mass distributions  
for $e^+ e^-$ candidate pairs where both tracks were identified as electrons.
We observe one event in the $B^0_s$ mass window, and two events 
in the $B^0$ mass window.  We estimate the total background contributions to be  $2.7 \pm 1.8$ events 
in both the $B^0_s$  and  $B^0$ mass windows. The dominant contribution comes from combinatorial background: $2.7 \pm 1.8$ 
compared to the contribution where both tracks are misidentified as electrons: $0.038\pm 0.008$ for both $B^0_s$  or  $B^0$.

We use the reference decay  $\bdkpdecay$ to set a limit on $\brsel$ (where $\ell$ is either $e$ or $\mu$), 
using the following expression:
\begin{eqnarray*}
\brsel  = & \frac {N(\bseldecay)\cdot  \brbdkp  \cdot f_d/f_s }{\epsilon^{rel}_{\bseldecay}   \cdot N( \bdkpdecay)}. 
\end{eqnarray*}
The expression for the $B^0$ channels is identical, except that the ratio of $b$--quark fragmentation 
probabilities: $f_d/f_s$ is not present.
In the expression, $N (\bseldecay)$ is the calculated upper limit on the number of $\bseldecay$ events, 
$N(\bdkpdecay)$ is the observed number of events from  the reference channel $\bdkpdecay$, 
$\brbdkp = (19.4\pm0.6)\times 10^{-6}$~\cite{PDG} is the branching ratio for  the $\bdkpdecay$ decay, 
and $\epsilon^{rel}_{\bseldecay}$ is  the detector acceptance and event selection  efficiency for 
reconstructing $\bseldecay$ decays relative to that for $\bdkpdecay$. 
The value of $f_d/f_s$  is 
$3.86 \pm 0.59$, 
where the (anti-)correlation between the uncertainties has 
been accounted for \cite{HFAG2006}.
To calculate the detector acceptance, we use simulated events with a 
detailed  simulation of the CDF II detector and event selection.  
We obtain  $\epsilon^{rel}_{\bsdecay} = 0.207 \pm  0.016$,  
$\epsilon^{rel}_{\bddecay} = 0.210 \pm 0.012$,  
$\epsilon^{rel}_{B^0_s \rightarrow e^+e^-} = 0.129  \pm 0.011$, and 
$\epsilon^{rel}_{B^0 \rightarrow e^+e^-} = 0.128  \pm 0.011$.
The uncertainties listed above  are the combined statistical and systematic uncertainties. The later include
uncertainties from  detector fiducial coverage, electron and muon identification 
efficiencies,  detector material determination,  $B^0_{(s)}$ $p_T$  spectrum,
 and  $B^0_{(s)}$ lifetimes.
The reference channel  $\bdkpdecay$ has been reconstructed using the same selection criteria except lepton 
identification.  
We find  $6387 \pm  214$ $\bdkpdecay$ events, using a fitting procedure similar to that described in Ref. \cite{b2hh-180pb}.

The upper limit on the branching ratio in each search window is  obtained 
using the Bayesian approach~\cite{PDG}, assuming a flat prior, and incorporating 
Gaussian uncertainties into the limit. The total systematic uncertainties, listed in Table~\ref{systotal}, are  
used as input for the limit calculation. 
Table~\ref{results} lists the upper limits  we obtain  
on the branching ratios at 90\% (95\%) credibility level (C.L.).

\begin{table*}[hptb]\centering
\caption{ Values used to calculate the limits on $\mathcal{B}(B^0_{(s)}\rightarrow e^+ \mu^-)$ and  $\mathcal{B}(B^0_{(s)}\rightarrow e^+ e^-)$ and their uncertainties.}
\begin{tabular}{lccccc}
\hline
\hline
Source & Values & $\mathcal{B}(B^0_s\rightarrow e^+ \mu^-)$ &
$\mathcal{B}(B^0\rightarrow e^+ \mu^-)$  & 
$\mathcal{B}(B^0_s\rightarrow e^+ e^-)$ &
$\mathcal{B}(B^0\rightarrow e^+ e^-)$ 
\\\hline 
$N(B^0\rightarrow K^+\pi^-)$ &$6387 \pm  214$&3.4\% &3.4\%&3.4\% &3.4\%\\
$\mathcal{B}(B^0\rightarrow K^+\pi^-)$ &$(19.4\pm0.6)\times 10^{-6}$&3.1\% & 3.1\%&3.1\% & 3.1\%\\
$f_{B^0}/f_{B^0_s}$ &$3.86\pm 0.59$ &15.3\%& - &15.3\%& - \\
$\epsilon^{rel}_{B^0_s\rightarrow e^+ \mu^- }$  & $0.207\pm 0.016$&7.6\%&-&-&-\\ 
$\epsilon^{rel}_{B^0\rightarrow e^+ \mu^-}$  & $0.210\pm 0.012$&-&5.9\%&-&-\\
$\epsilon^{rel}_{B^0_s\rightarrow e^+ e^- }$  & $0.129\pm 0.011$&-&-&8.9\%&-\\ 
$\epsilon^{rel}_{B^0\rightarrow e^+ e^-}$  & $0.128\pm 0.011$&-&-&-&8.9\%\\
\hline
Total&&17.7\% & 7.5\% &18.3\%&10.0\%\\
\hline
\hline
\end{tabular}
\label{systotal}
\end{table*}

\begin{table}[hptb]\centering
\caption{Branching ratio limits at 90(95)~\% C.L.}
\begin{tabular}{c}
\hline
\hline
$\brslimit$\\
$\brdlimit$\\
$\brstoeelimit$\\ 
$\brdtoeelimit$\\
\hline
\hline
\end{tabular}
\label{results}
\end{table}

\begin{figure}[htbp]
\begin{center}
\includegraphics[width=3.5in]{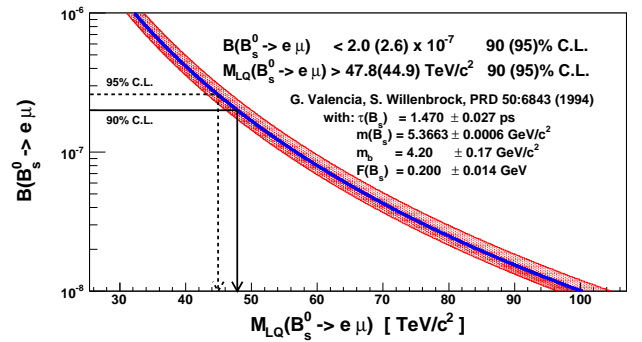}
\caption{ Leptoquark mass limit corresponding to the  90 (95)  \% C.L. on
 $\mathcal{B}(B^0_s\rightarrow e^+ \mu^-)$. The error band is obtained by varying the values entering 
the theoretical calculation  within their uncertainties. The uncertainties stemming from approximating
 $\alpha_s$ are not included.  
}
\label{lqlimit}
\end{center}
\end{figure}

Within the Pati-Salam leptoquark  model, the following relationship 
between the $\mathcal{B}(B^0_{(s)}\rightarrow e^+ \mu^-)$ and 
the leptoquark mass ($M_{LQ}$) can be derived~\cite{scott}:

\begin{eqnarray*}
\mathcal{B}(B^0_{(s)}\rightarrow e^+ \mu^-)
&=& \pi\alpha^2_s(M_{LQ}) \frac{1}{M^4_{LQ}} F^2_{B^0_{(s)}} m^3_{B^0_{(s)}} R^2    \cdot \frac{\tau_{B^0_{(s)}}}{\hbar},\\
\end{eqnarray*}
\noindent 
where $R= \frac{m_{B^0_{(s)}}}{m_b}\left(\frac{\alpha_s(M_{LQ})}{\alpha_s(m_t)}\right)^{-\frac{4}{7}}
\left(\frac{\alpha_s(m_t)}{\alpha_s(m_b)}\right)^{-\frac{12}{23}}$.
The values and uncertainties of the quantities used
in the calculation of $M_{LQ}$ are the following \cite{PDG}:
the top-quark mass  $m_t$ (171.2  $\pm$ 2.1  GeV/c$^2$),
the bottom quark mass    $m_b$ (4.20  $\pm$ 0.17 GeV/c$^2$),
the charm quark mass    $m_c$ (1.27  $\pm$ 0.11 GeV/c$^2$), 
the $B^0$-meson mass $m_{B^0}$         (5.27953 $\pm$ 0.00033 GeV/c$^2$),
the $B^0_s$-meson mass $m_{B^0_s}$         (5.3663 $\pm$ 0.0006 GeV/c$^2$),
the $B^0$-meson lifetime $\tau_{B^0}$  (1.530 $\pm$ 0.009 ps),  
the $B^0_s$-meson lifetime $\tau_{B^0_s}$  (1.470 $\pm$ 0.027 ps),
the coupling strength  $F_{B^0}$  (0.178  $\pm$ 0.014 GeV), 
and $F_{B^0_s}$  (0.200  $\pm$ 0.014 GeV)\cite{constant}. 
For the strong coupling constant we use  $\alpha_s(M_{Z^0})$ = 0.115, which is evolved to  $M_{LQ}$ using the Marciano 
approximation~\cite{marciano} assuming no colored particles exist with masses  between  $m_t$ and $M_{LQ}$.   
Using the limits 
on the branching ratios listed in Table~\ref{results}, we calculate limits on the masses of the corresponding
Pati-Salam leptoquarks of 
$ \psslimit$ 
\noindent
and $\psdlimit$ at 90 (95)\% C.L. 
Figure~\ref{lqlimit} shows the limit and the relation between the
leptoquark mass and the branching ratio for the $B^0_s$ meson.

In summary, we report on a search for the lepton flavor violating decays  $B^0_{(s)}\rightarrow e^+\mu^-$ 
and the flavor changing neutral current decays $B^0_{(s)} \rightarrow e^+ e^-$ 
using data corresponding to 2 $\fb$ of integrated luminosity collected in 
$p\overline{p}$ collisions at $\sqrt{s} = 1.96$~TeV.
This is the first search for $B^0_s  \rightarrow e^+ e^-$ decays.
We observe no evidence for these decays and set limits that are the most stringent to date. 
These results represent 
a significant improvement compared to the previous measurement~\cite{CDF_RUNI} by CDF and the 
best results from $B$-Factories~\cite{babar07,belle03,CLEO2}.  

We thank the Fermilab staff and the technical staffs of the participating institutions for their vital contributions.
This work was supported by the U.S. Department of Energy and National Science Foundation; 
the Italian Istituto Nazionale di Fisica Nucleare;
 the Ministry of Education, Culture, Sports, Science and Technology of Japan; 
the Natural Sciences and Engineering Research Council of Canada; the National Science Council of the Republic of China;
 the Swiss National Science Foundation; the A.P. Sloan Foundation;
 the Bundesministerium f\"ur Bildung und Forschung, Germany;
 the Korean Science and Engineering Foundation and the Korean Research Foundation;
 the Science and Technology Facilities Council and the Royal Society, UK; 
the Institut National de Physique Nucleaire et Physique des Particules/CNRS;
 the Russian Foundation for Basic Research;
 the Ministerio de Ciencia e Innovaci\'{o}n, and Programa Consolider-Ingenio 2010, Spain; the Slovak R\&D Agency;
 and the Academy of Finland.



\begin{thebibliography}{99.}
\bibitem{qconj}
 Throughout this Letter inclusion of charge conjugate reactions is implied.
\bibitem{csys}
The polar angle $(\theta)$ in cylindrical coordinates is  measured with 
respect to the proton beam direction, which defines the $z$-axis. 
Pseudorapidity $(\eta)$ is defined as 
$\eta = -\ln(\tan\frac{\theta}{2})$.
\bibitem{IP}
The impact parameter $d_0$ is the distance of closest approach of the track to the beam line.
\bibitem{isodef} Due to the hard $b$-quark fragmentation, $B$-mesons carry most of the momentum of the
$b$-quark. 
The isolation is defined as $\iso= \ptB  /(\sum_{i}p^{i}_{T} +  \ptB  )$,
where $\ptB$ is the transverse momentum of the $B$ candidate, and the sum runs over all other tracks 
within a cone of radius 1 in $\eta-\phi$ space around the $B$  flight direction.
\bibitem{poidef}
For track pairs coming from the two-body decay of a $B$, the vector pointing from the 
primary vertex to the $B$ decay vertex in the transverse plane  $\vec{l}_{xy}$  
should point in the same direction as the transverse momentum vector $\vec{p}_{\rm{T}}$(B) of the $B$ candidate. 
$\Delta \phi$ is defined as the angle  between $\vec{l}_{xy}$ and  $\vec{p}_{\rm{T}} (B)$.
\end{thebibliography}

\end{document}